\documentclass[11pt]{article}

\usepackage[a4paper,margin=2.5cm]{geometry}

\usepackage{amsmath,amssymb,amsfonts}

\usepackage{authblk}

\usepackage{booktabs,longtable,array,multirow,makecell}

\usepackage{graphicx}
\usepackage{float}
\usepackage{adjustbox}

\usepackage{cite}

\usepackage[colorlinks=true, linkcolor=blue, urlcolor=blue, citecolor=blue]{hyperref}

\usepackage{caption}
\usepackage{setspace}
\usepackage{enumitem}
\usepackage{titlesec}

\usepackage{indentfirst}

\usepackage[scheme=plain]{ctex}

\usepackage{etoolbox}

\makeatletter
\renewenvironment{abstract}
 {
  \small
  \begin{center}
    {\bfseries Abstract}
  \end{center}
  \parindent=2em
 }
 {}
\makeatother

\titleformat{\section}
{\large\bfseries}
{\thesection.}
{0.5em}
{}

\titleformat{\subsection}
{\normalsize\bfseries}
{\thesubsection.}
{0.5em}
{}

\graphicspath{{../figures/}{figures/}}

\newcommand{\mpl}{M_{\rm Pl}}
\newcommand{\dd}{\mathrm{d}}
\newcommand{\PR}{\mathcal P_{\mathcal R}}
\newcommand{\PT}{\mathcal P_T}
\newcommand{\Geff}{Z_{\rm eff}^{(A_s,n_s,r)}}

\title{\LARGE Prior-Averaged Ranking of Low-Order Monomial Potentials in Low-Temperature Warm Inflation}

\author[1]{Xin Peng}
\author[2]{Xiang Cheng}
\author[1]{Wei Cheng\thanks{Corresponding authors. Email: chengwei@cqupt.edu.cn, mayr@cqupt.edu.cn, panyu@cqupt.edu.cn, zengj@hainnu.edu.cn}}
\author[1]{Yi-Rong Ma$^{*}$}
\author[1]{Yu Pan$^{*}$}
\author[3]{Jun Zeng$^{*}$}

\affil[1]{School of Electronic Science and Engineering, Chongqing University of Posts and Telecommunications, Chongqing 400065, China}
\affil[2]{School of Communication and Information Engineering, Chongqing University of Posts and Telecommunications, Chongqing 400065, China}
\affil[3]{College of Physics and Electronic Engineering,
Hainan Normal University, Haikou 571158, Hainan, China}

\date{}

\begin{document}

\maketitle

\vspace{0.5em}
\begin{abstract}
We compare the relative prior-averaged weights of the monomial potentials
\(V_p(\phi)=\lambda_p\phi^p/p\), with \(p=2,3,4\), in low-temperature warm inflation with the dissipative coefficient fixed to
\(\Upsilon=C_\phi T^3/\phi^2\). The dissipative form, background equations, primordial-spectrum prescription, and compressed observable likelihood are held fixed while the monomial power is varied. For each branch, the warm background equations, including radiation backreaction, are solved numerically, and the broadened likelihood for \((A_s,n_s,r_{0.05})\) is integrated over the adopted prior domain to obtain \(Z_{\rm eff}^{(A_s,n_s,r)}\). For the reference configuration \(N_*=55\), \(\sigma_r=0.005\), and the viability-conditioned prior A, we obtain
\(\Delta\ln Z_{\rm eff}(p=2)=-32.18\) and \(\Delta\ln Z_{\rm eff}(p=3)=-6.99\) relative to \(p=4\). The same ranking is retained for the specific variations of \(N_*\), the prior domains, the random seeds, and the smoothing width of the upper-bound penalty examined here. A representative quartic trajectory gives \(n_s=0.96420\), \(r_{0.05}=0.02663\), \(Q_*=4.68\times10^{-3}\), and \(T_*/H_*=10.67\). Its scalar spectrum is evaluated using the assumed Bose--Einstein occupation prescription. A component test at this representative point indicates that the occupation term contributes more to the scalar enhancement than the fitted dissipative growth factor. Thus, within the compressed-likelihood setup, the adopted thermal-occupation prescription, and the adopted viability-conditioned priors, the prior-averaged ranking is \(p=4>p=3\gg p=2\).

\end{abstract}

\clearpage
\begin{center}
{\Large\bfseries Contents}
\end{center}
\vspace{0.6em}
\begingroup
\makeatletter
\@starttoc{toc}
\makeatother
\endgroup
\clearpage

\section{Introduction}\label{sec:intro}

Inflation provides a dynamical explanation for the origin of the large scale homogeneity, flatness, and nearly scale invariant curvature perturbations of the universe \cite{Guth:1980zm, Linde:1981mu, Albrecht:1982wi, Lyth:1998xn, Baumann:2009ds}. In the minimal cold single-field framework, the inflaton is assumed to remain approximately isolated from other degrees of freedom during the accelerated expansion phase, while scalar perturbations are generated predominantly from quantum fluctuations on a quasi-de Sitter background \cite{Tasinato:2020vdk, Adshead:2017srh, Yuan:2026xcg, Baumann:2007zm}. The corresponding observational predictions are conventionally characterized by the scalar amplitude \((A_s)\), the scalar spectral index \((n_s)\), and the tensor-to-scalar ratio \((r)\). Constraints from the Planck 2018 inflation analysis yield \(n_s=0.9649\pm0.0042\) (68\% CL), while the commonly adopted bound from the BICEP/Keck 2018 analysis gives \(r_{0.05}<0.036\) (95\% CL) \cite{Planck:2018jri, BICEP:2021xfz, Planck:2018nkj, Sheikhahmadi:2019gzs}.

Monomial potentials \(V(\phi)\propto\phi^p\), originally motivated by chaotic inflation and subsequent effective field theory considerations, continue to serve as important benchmark models in inflationary cosmology \cite{Linde:1983gd, Harigaya:2014wta, Rezazadeh:2015dza, Shahalam:2017wba, Ballesteros:2015noa}. Simple analytic relations are obtained among the power \(p\), the pivot e-fold number \(N_*\), \(n_s\), and \(r\), while these models may also be interpreted as low-order operators in an effective scalar potential. Owing to this simplicity, the corresponding observational constraints become particularly transparent. In cold slow-roll inflation, \(r\) increases with the monomial power \(p\); consequently, for \(N_*\simeq50\text{--}60\), the quartic potential generally predicts values of \(r\) significantly above the current observational upper bound. This difficulty does not correspond merely to a single point fitting issue, but instead reflects a structural limitation arising from the direct connection between the tensor amplitude and the slope of the potential within the cold inflationary background.

Warm inflation modifies these relations by allowing the inflaton to dissipate energy into a light thermal bath during inflation \cite{Bartrum:2013fia, Berera:2008ar, Berera:2023liv, Kamali:2023lzq}. At the background level, the term \(\Upsilon\dot{\phi}\) adds friction and sources radiation. At the perturbative level, the scalar spectrum can receive contributions from thermal noise, a non-vacuum inflaton occupation state, and the coupled inflaton--radiation system \cite{Hall:2003zp, Graham:2009bf, Ramos:2013nsa, Montefalcone:2023pvh}. Because the tensor spectrum is not enhanced by the same terms, the ratio \(r=\PT/\PR\) can be lower than in the corresponding cold model. Previous analyses have shown that this mechanism can move quartic or plateau-type predictions toward regions allowed by CMB constraints, including in warm Higgs and Palatini/\(R^2\) settings \cite{Eadkhong:2023ozb, Yuennan:2025szw, Cheng:2024uvn}. These results provide part of the established context for the comparison undertaken below.

The low-temperature coefficient used here has an established effective-field-theory interpretation. In the commonly studied two-stage mechanism, \(\Upsilon(\phi,T)=C_\phi T^3/\phi^2\) applies when the bath temperature lies below the heavy-mediator scale \cite{Moss:2006gt, Bastero-Gil:2006ahd, Bastero-Gil:2009sdq, Bastero-Gil:2012akf}. The low-temperature condition is \(T<m_\chi(\phi)\), whereas \(T/H>1\) characterizes a warm thermal background. A separate rate condition is required for equilibration within the light bath, and neither condition alone establishes a Bose--Einstein distribution for the inflaton fluctuations. In the numerical analysis, \(T/H\), \(T/\phi\), \(Q\), and \(\rho_R/V\) are therefore used to diagnose the background regime, while the inflaton occupation state is stated explicitly as an additional perturbative assumption.

The ingredients entering the present calculation have been studied extensively. Low-temperature two-stage dissipation, distributed-mass constructions, the Warm Little Inflaton scenario, Minimal Warm Inflation, and gauge-interaction realizations provide established microphysical settings \cite{Bastero-Gil:2009sdq, Bastero-Gil:2012akf, Bastero-Gil:2016qru, Bastero-Gil:2018yen, Berghaus:2019whh, Berghaus:2025dqi}. CMB constraints on warm inflation have been examined with several statistical treatments \cite{Bartrum:2013oka, Benetti:2016jhf, Bastero-Gil:2017wwl, Arya:2017zlb}. Monomial potentials and the observational role of a thermal inflaton occupation have likewise been discussed in detail \cite{Bartrum:2013fia, Visinelli:2016rhn, Ballesteros:2023dno}. In addition, the quartic potential has been analyzed for several dissipative coefficients, including \(\Upsilon\propto T^3/\phi^2\) \cite{Santos:2024pix}. The perturbation formula used below also relies on analytic and numerical results developed in earlier work \cite{Hall:2003zp, Graham:2009bf, Ramos:2013nsa, Montefalcone:2023pvh, Laine:2025rll}.

No claim is made here that the low-temperature coefficient, the quartic warm-inflation branch, the thermal occupation contribution, or the use of CMB-based statistical inference is new. The question considered here is narrower. We fix \(\Upsilon=C_\phi T^3/\phi^2\), the background equations, the spectrum prescription, and the compressed observable likelihood, and vary only the monomial power \(p=2,3,4\). This is complementary to studies that fix the quartic potential and compare several dissipative coefficients. The purpose is to determine the prior-averaged ordering obtained for these three branches within one numerical setup and a specified family of viability-conditioned priors.

The three potentials \(V_p(\phi)=\lambda_p\phi^p/p\), with \(p=2,3,4\), are evaluated in the same low-temperature setup. The cubic case is treated only as a local effective branch for \(\phi>0\). For each power, representative high-likelihood trajectories are reported together with the prior integral \(Z_{\rm eff}\) constructed from the broadened likelihood in \((A_s,n_s,r_{0.05})\). Variations of \(N_*\), the prior domains, random seeds, the upper-bound smoothing width, and selected spectrum settings are adopted to examine whether the ranking changes within the configurations tested.

All three branches are evolved with identical background equations and the same primordial-spectrum prescription. Prior A is defined after exploratory viability scans, and priors B and C probe finite changes in the adopted domains. The background checks include \(T<m_\chi\), \(T/H>1\), radiation subdominance during the CMB window, and a light-bath equilibration estimate. These checks do not by themselves derive the inflaton occupation number. The baseline spectrum instead adopts a Bose--Einstein occupation as an explicit assumption. Setting \(n_*=0\) is used only as a limiting component test at the representative trajectories and is not a new evidence integration under an alternative perturbation model.

The remainder of the paper is organized as follows. Sec.~\ref{sec:model} introduces the monomial potentials, the cold-inflation benchmark, and the low-temperature effective description. Sec.~\ref{sec:warm} presents the background equations, the spectrum prescription, the observable calculation, and the compressed likelihood. Sec.~\ref{sec:statistics} describes the parameterization, the viability-conditioned priors, and the sampling procedures. Sec.~\ref{sec:results} reports the representative trajectories, the conditional \(Z_{\rm eff}\) ordering, the sensitivity tests, and the background diagnostics. Sec.~\ref{sec:discussion} discusses the occupation assumption, the interpretation of \(Z_{\rm eff}\), the prior dependence, and the microphysical cost. Sec.~\ref{sec:conclusion} summarizes the scope and limitations of the result.

\section{Model Setup and Cold-Inflation Benchmark}\label{sec:model}
\subsection{Low-Order Monomial Potentials}

Throughout the paper, the potential is written as
\begin{equation}
  V_p(\phi)=\frac{\lambda_p}{p}\phi^p,\qquad p=2,3,4 .
  \label{eq:potential-body}
\end{equation}
In reduced Planck units, \(V\) has mass dimension four and \(\lambda_p\) has mass dimension \(4-p\). The cases \(p=2\) and \(p=4\) are the usual quadratic and quartic terms. The \(p=3\) case is retained only as a local effective branch over the positive-field interval relevant to the CMB calculation. A globally stable completion would require additional operators or a specified symmetry-breaking sector. The conclusions for \(p=3\) therefore concern this local branch and not an isolated cubic potential over the full field domain.

\subsection{Analytic Cold-Inflation Benchmark}

For \(V\propto\phi^p\), the cold slow-roll parameters are
\begin{equation}
\epsilon_V=\frac12\left(\frac{V_{,\phi}}{V}\right)^2
=\frac{p^2}{2\phi^2},\qquad
\eta_V=\frac{V_{,\phi\phi}}{V}
=\frac{p(p-1)}{\phi^2}.
\end{equation}
where \(V_{,\phi}\equiv \mathrm{d}V/\mathrm{d}\phi\) and \(V_{,\phi\phi}\equiv \mathrm{d}^2V/\mathrm{d}\phi^2\). The ratios \(V_{,\phi}/V\) and \(V_{,\phi\phi}/V\) measure the relative slope and curvature of the potential and enter the slow-roll conditions. The comma notation is used throughout for derivatives with respect to \(\phi\).

The condition for the end of cold slow-roll inflation is
\(\epsilon_V(\phi_{\rm end})=1\), hence
\(\phi_{\rm end}=p/\sqrt2\). The pivot field value satisfies
\begin{equation}
N_*\simeq\int_{\phi_{\rm end}}^{\phi_*}\frac{V}{V_{,\phi}}\dd\phi
=\frac{\phi_*^2-\phi_{\rm end}^2}{2p},\qquad
\phi_*^2=2pN_*+\frac{p^2}{2}.
\end{equation}
Here and throughout the paper, the subscript $*$ denotes quantities evaluated at horizon crossing of the pivot-scale \(k_*=0.05\,\mathrm{Mpc}^{-1}\). Hence, the cold slow-roll parameters can be derived
\begin{equation}
\epsilon_{V,*}=\frac{p}{4N_*+p},\qquad
\eta_{V,*}=\frac{2(p-1)}{4N_*+p}.
\end{equation}

The first order slow-roll relations in cold inflation are
\begin{equation}
 n_s^{\rm cold}-1=-6\epsilon_{V,*}+2\eta_{V,*},\qquad
 r^{\rm cold}=16\epsilon_{V,*}.
 \label{eq:cold-sr-relation}
\end{equation}

Substituting the slow-roll parameters yields
\begin{equation}
 n_s^{\rm cold}=1-\frac{2(p+2)}{4N_*+p},\qquad
 r^{\rm cold}=\frac{16p}{4N_*+p}.
 \label{eq:cold-nsr}
\end{equation}
 
For \(N_*=55\), the cold inflation benchmark results are listed in Tab.~\ref{tab:cold}. These results are included solely for reference purposes and are not incorporated into the effective evidence ranking of the warm inflation models.

\begin{table}[htbp]
\centering
\caption{Analytic benchmark for cold monomial potentials with \(N_*=55\). The last column gives the qualitative assessment relative to the current \(n_s\) and \(r_{0.05}\) constraints.}
\label{tab:cold}
\resizebox{\textwidth}{!}{%
\footnotesize
\begin{tabular}{ccccccp{5.4cm}}
\toprule
Model & \(N_*\) & \(\phi_*\) & \(\epsilon_{V,*}\) & \(n_s^{\rm cold}\) & \(r^{\rm cold}\) & \multicolumn{1}{c}{Qualitative interpretation} \\
\midrule
\(p=2\) & 55 & 14.8997 & 0.0090 & 0.9640 & 0.1441 &  \(n_s\) is close to the observational central values, but \(r\) is clearly too large. \\
\(p=3\) & 55 & 18.2893 & 0.0135 & 0.9552 & 0.2152 &  \(n_s\) is on the low side, and \(r\) is even larger than for the quadratic potential. \\
\(p=4\) & 55 & 21.1660 & 0.0179 & 0.9464 & 0.2857 &  \(n_s\) is further on the low side, and \(r\) is the largest. \\
\bottomrule
\end{tabular}%
}
\end{table}

\subsection{Low-Temperature Dissipation}
The analysis adopts the low-temperature dissipative form
\begin{equation}
  \Upsilon(\phi,T)=C_\phi\frac{T^3}{\phi^2}.
  \label{eq:dissipation}
\end{equation}
Here \(C_\phi\) is a dimensionless effective coefficient that may contain multiplicity and coupling factors from the underlying two-stage dissipative sector \cite{Moss:2006gt, Bastero-Gil:2006ahd, Bastero-Gil:2012akf}. If the heavy-mediator mass is parameterized as \(m_\chi(\phi)=g_\chi\phi\), the low-temperature expansion requires
\begin{equation}
T<m_\chi,
\qquad \hbox{or equivalently}\qquad
g_\chi>\frac{T}{\phi}.
\label{eq:gchi-condition}
\end{equation}

If the equilibration rate within the light radiation sector is parameterized as \(\Gamma_{\rm th}=\kappa T\), the corresponding estimate is
\begin{equation}
\frac{\Gamma_{\rm th}}{H}>1,
\qquad
\kappa>\frac{H}{T}.
\label{eq:kappa-condition}
\end{equation}
This condition concerns equilibration of the light bath. It does not by itself imply that inflaton perturbations have a Bose--Einstein occupation number.

The inequalities are evaluated at the pivot and over the reported e-fold intervals. Unless stated otherwise, \(g_*=228.75\) is used as a benchmark value for the relativistic content of the bath. This choice does not specify a unique particle model. The quantities \(g_*\) and \(C_\phi\) have different roles: \(g_*\) enters the relation between \(\rho_R\) and \(T\), whereas \(C_\phi\) parameterizes the effective dissipative channel. No one-to-one relation between them is assumed.

Using
\begin{equation}
  \frac{m_\chi}{H}=g_\chi\frac{\phi}{H}
  =g_\chi\frac{T/H}{T/\phi},
  \label{eq:mchiH-check}
\end{equation}
the representative CMB windows, for which \(T/H\sim10\) and \(T/\phi\sim10^{-5}\), correspond to \(\phi/H\sim10^6\). The inequalities \(T<m_\chi\) and \(m_\chi\gg H\) can therefore be satisfied for a range of \(g_\chi\). This check supports use of the heavy-mediator effective description along the reported trajectories, but it does not supply a microscopic construction. In particular, the required \(C_\phi\) is large and remains a model-building requirement of the present effective treatment.

\section{Warm Background, Primordial Spectrum, and Compressed Likelihood}\label{sec:warm}

\subsection{Warm-Inflation Background Equations}\label{sec:bg-eq}

The homogeneous dynamics are described by
\begin{gather}
\ddot\phi+(3H+\Upsilon)\dot\phi+V_{,\phi}=0,
\label{eq:eomphi}\\
\dot\rho_R+4H\rho_R=\Upsilon\dot\phi^2,
\label{eq:eomrho}\\
H^2=\frac{1}{3\mpl^2}
\left(\frac12\dot\phi^2+V+\rho_R\right).
\label{eq:friedmann}
\end{gather}
Here \(H=\dot a/a\), and an overdot denotes a derivative with respect to cosmic time. The term \(\Upsilon\dot\phi\) transfers energy from the inflaton to radiation, while Eq.~\eqref{eq:friedmann} constrains the total background energy density.

The radiation energy density is related to the temperature through
\begin{equation}
\rho_R=\frac{\pi^2}{30}g_*T^4.
\label{eq:rhoT}
\end{equation}

The principal calculations use \(g_*=228.75\). This benchmark fixes the conversion between \(T\) and \(\rho_R\) through Eq.~\eqref{eq:rhoT}; it does not specify the particle content of a unique microscopic model. The coefficient \(C_\phi\) is treated independently as an effective parameter in \(\Upsilon=C_\phi T^3/\phi^2\), so that \(\Upsilon\) changes along the background trajectory with \(\phi\) and \(\rho_R\).

The dissipation ratio is
\begin{equation}
Q\equiv\frac{\Upsilon}{3H}.
\label{eq:Qdef}
\end{equation}
Values \(Q\ll1\) correspond to a background dominated by Hubble friction, whereas \(Q\gtrsim1\) indicates a substantial dissipative contribution. The representative CMB trajectories reported in Sec.~\ref{sec:results} remain in the weak-dissipation regime; \(Q\) can grow only during the final radiation-enhanced stage.

The Hubble slow-roll parameter may be expressed directly from energy conservation as
\begin{equation}
\dot H=-\frac{1}{2\mpl^2}\left(\dot\phi^2+\frac{4}{3}\rho_R\right),
\qquad
\epsilon_H\equiv-\frac{\dot H}{H^2}
=\frac{\dot\phi^2+4\rho_R/3}{2\mpl^2H^2} .
\label{eq:epsilonH}
\end{equation}

Consequently, the condition for accelerated expansion is given by \(\epsilon_H<1\), while the end of inflation is defined by \(\epsilon_H=1\). This criterion explicitly incorporates the contribution of the radiation energy density to \(\dot H\), and is therefore more appropriate for warm inflationary backgrounds than a description based solely on the potential slow-roll parameter.

In the numerical analysis, the e-fold number \(N\equiv\ln a\) is adopted as the time variable. By defining \(\pi_\phi\equiv\dot{\phi}\) and using the relation \(\dd/\dd t=H\dd/\dd N\), Eqs.~\eqref{eq:eomphi}--\eqref{eq:eomrho} may be rewritten in first order form as

\begin{align}
\frac{\dd\phi}{\dd N}&=\frac{\pi_\phi}{H},\label{eq:dphidN}\\
\frac{\dd\pi_\phi}{\dd N}&=-\frac{(3H+\Upsilon)\pi_\phi+V_{,\phi}}{H},\label{eq:dpidN}\\
\frac{\dd\ln\rho_R}{\dd N}&=-4+\frac{\Upsilon\pi_\phi^2}{H\rho_R} .\label{eq:drhodN}
\end{align}

The third equation is written in terms of \(\ln\rho_R\) in order to preserve the positivity of the radiation energy density and to improve numerical stability when \(\rho_R\) spans multiple orders of magnitude. At each integration step, the temperature \(T\) is determined from Eq.~\eqref{eq:rhoT}, the dissipative coefficient \(\Upsilon\) is obtained from Eq.~\eqref{eq:dissipation}, and the Hubble expansion rate \(H\) is evaluated through the Friedmann Eq.~\eqref{eq:friedmann}, thereby yielding a closed system for the background evolution.

In the slow-roll regime with subdominant radiation energy density, the full background equations reduce to the standard warm inflation slow-roll relations:
\begin{equation}
3H(1+Q)\dot\phi\simeq -V_{,\phi},
\qquad
4H\rho_R\simeq \Upsilon\dot\phi^2,
\qquad
H^2\simeq \frac{V}{3\mpl^2} .
\label{eq:warm-sr}
\end{equation}

These relations provide a useful physical interpretation of warm inflation: the dissipative friction suppresses \(|\dot{\phi}|\) through the factor \(1+Q\), while the source term \(\Upsilon\dot{\phi}^2\) continuously maintains the thermal radiation bath. Nevertheless, the numerical results presented here are not based on the analytic approximation of Eq.~\eqref{eq:warm-sr}, but are instead obtained through direct numerical integration of the background Eqs.~\eqref{eq:dphidN}--\eqref{eq:drhodN}. The location of the pivot-scale is specified by the reference condition \(N_{\rm end}-N_*=55\), and the stability of the results is further examined for \(N_*=50,55,60\).

\subsection{Primordial Spectrum and Occupation Assumption}
\label{sec:primordial-spectra}

The scalar spectrum is evaluated with the standard fitted prescription
\cite{Hall:2003zp, Graham:2009bf, Ramos:2013nsa, Montefalcone:2023pvh}
\begin{equation}
\PR(k_*)=
\left(\frac{H_*^2}{2\pi |\dot\phi_*|}\right)^2
\left[
1+2n_*+
\frac{2\sqrt3\pi Q_*}{\sqrt{3+4\pi Q_*}}\frac{T_*}{H_*}
\right]G(Q_*).
\label{eq:PR}
\end{equation}
The prefactor is the vacuum contribution. The bracket contains the assumed initial occupation term and the thermal-noise contribution, while \(G(Q)\) represents the fitted growth of the coupled inflaton--radiation perturbations. As a baseline prescription, the occupation number is taken to be
\begin{equation}
 n_* = \frac{1}{\exp(H_*/T_*)-1}.
\label{eq:nBE}
\end{equation}
Equation~\eqref{eq:nBE} is an assumption about the inflaton fluctuations and is not derived from the background condition \(T_*/H_*>1\) or from equilibration of the light bath.

Under the adopted Bose--Einstein prescription, \(T_*/H_*>1\) gives a non-negligible occupation term. In the limit \(T_*/H_*\gg1\),
\begin{equation}
 n_* \simeq \frac{T_*}{H_*},
 \qquad
 1+2n_*\simeq 1+2\frac{T_*}{H_*}.
 \label{eq:thermal-occupation-estimate}
\end{equation}
For the representative quartic trajectory, \(T_*/H_*=10.67\), so this term increases the scalar amplitude and lowers \(r=\mathcal P_T/\mathcal P_{\mathcal R}\) relative to the same trajectory evaluated with \(n_*=0\).

The baseline values of \(Z_{\rm eff}\) are conditional on Eq.~\eqref{eq:nBE}. The replacement \(n_*=0\) used later is applied only to the representative trajectories to isolate the size of the occupation contribution. It is not a new nested-sampling calculation and does not determine the \(Z_{\rm eff}\) ordering for a nonthermal or partially thermal inflaton state.

The function \(G(Q)\) characterizes the growth effect induced by the coupling between inflaton and radiation perturbations. It is not an additional free function, but an empirical fitting factor extracted from numerical solutions of the warm inflation perturbation equations. This type of fitting factor has been widely used in analyses of warm inflation perturbation spectra and CMB constraints
\cite{Kamali:2023lzq, Montefalcone:2023pvh, Bastero-Gil:2016qru, Benetti:2016jhf}.
For a dissipative coefficient with cubic temperature dependence, a commonly used fitting form is
\begin{equation}
G_{\mathrm{cubic}}(Q_*)=
1+4.981\,Q_*^{1.946}+0.127\,Q_*^{4.330}.
\label{eq:GQ}
\end{equation}
This fitting expression summarizes the numerical growth effect from the coupled perturbation equations in the case \(\Upsilon\propto T^3\)
\cite{Montefalcone:2023pvh, Ito:2025lcg, Santos:2022exm}. The fitting coefficients depend on the temperature dependence of the dissipative coefficient. For example, in the case of linear temperature dependence, a commonly adopted fitting form is
\cite{Kamali:2023lzq}
\begin{equation}
G_{\mathrm{linear}}(Q_*)=
1+0.335\,Q_*^{1.364}+0.0185\,Q_*^{2.315}.
\label{eq:GQlinear}
\end{equation}

Because the coefficient in Eq.~\eqref{eq:dissipation} has cubic temperature dependence, Eq.~\eqref{eq:GQ} is used as the baseline fit. Along the representative CMB trajectories, \(Q\) is of order \(10^{-3}\), and the fitted value of \(G(Q)\) remains close to unity. The later variations of \(G(Q)\) and the replacement \(n_*=0\) are therefore used as component tests at those trajectories. They do not constitute alternative evidence calculations.

The tensor spectrum is taken to be
\begin{equation}
\PT(k_*)=
\frac{2H_*^2}{\pi^2\mpl^2},
\qquad
r_{0.05}=
\frac{\PT(k_*)}{\PR(k_*)}.
\label{eq:tensor}
\end{equation}
Within this prescription, changes in \(r\) arise mainly through the scalar denominator rather than a direct thermal modification of \(\PT\). The amplitude and tilt are evaluated on the same numerical trajectory as described in Sec.~\ref{sec:observable-flow}.

\subsection{Calculation of the Pivot Observables}
\label{sec:observable-flow}

For a given monomial power \(p\) and parameter set \(\theta\), the observables are evaluated according to the following procedure. The background system defined by Eqs.~\eqref{eq:eomphi}--\eqref{eq:friedmann} is first solved numerically, while the end of inflation is determined through the condition \(\epsilon_H=1\). The pivot point associated with a specified value of \(N_*\) is subsequently identified by moving backward from the end of inflation. At this pivot-scale, the quantities \(H_*\), \(\dot{\phi}_*\), \(Q_*\), \(T_*\), and \(\rho_{R,*}\) are extracted from the numerical trajectory. The scalar amplitude is then defined as
\begin{equation}
 A_s = \PR(k_*),
 \label{eq:As-def}
\end{equation}
where \(\PR\) is given by Eq.~\eqref{eq:PR}. The scalar spectral index is defined by
\begin{equation}
 n_s-1=
 \left.
 \frac{\dd\ln \PR}{\dd\ln k}
 \right|_{k_*}.
 \label{eq:ns-def}
\end{equation}
Using the horizon-crossing relation \(k=aH\), one has
\begin{equation}
 \dd\ln k=\dd\ln(aH)=(1-\epsilon_H)\dd N ,
 \label{eq:dlnk-dN}
\end{equation}
where \(N\equiv\ln a\) denotes the forward e-fold integration variable. To avoid ambiguity between this forward variable and the remaining number of e-folds before the end of inflation, the forward pivot position is defined as
\begin{equation}
 N_{\rm piv}=N_{\rm end}-N_*,
 \qquad
 N_\pm=N_{\rm piv}\pm\Delta N ,
 \label{eq:Npiv-def}
\end{equation}
where the quantity \(N_*\) appearing in the results tables represents the number of remaining e-folds between the pivot-scale and the end of inflation.

In the numerical implementation, \(\ln\PR\) is evaluated along the same background trajectory at \(N_-\), \(N_{\rm piv}\), and \(N_+\), with \(\Delta N=0.20\). These three values are obtained by shifting the horizon-crossing position along the forward variable \(N=\ln a\), reevaluating \(H\), \(\dot{\phi}\), \(Q\), \(T\), and \(\rho_R\), and substituting these quantities into Eq.~\eqref{eq:PR}. A local quadratic fit of \(\ln\PR\) as a function of \(N-N_{\rm piv}\) is then used to extract
\begin{equation}
 \left.
 \frac{\dd\ln\PR}{\dd N}
 \right|_* .
 \label{eq:dlnPR-dN-local}
\end{equation}
The scalar spectral index is finally computed through the local horizon-crossing conversion:
\begin{equation}
 n_s-1
 \simeq
 \frac{1}{1-\epsilon_{H,*}}
 \left.
 \frac{\dd\ln\PR}{\dd N}
 \right|_* .
 \label{eq:ns-finite-difference}
\end{equation}
Equation~\eqref{eq:ns-finite-difference} is the numerical procedure adopted in this work. It keeps the leading correction from the variation of \(H\) through the factor \(1-\epsilon_{H,*}\), rather than approximating \(\dd\ln k\) simply by \(\dd N\). The quantities \(\PR(N_\pm)\) are therefore not independently adjusted amplitudes; they are computed from the same warm-inflation trajectory by shifting the pivot location and reevaluating the local background quantities. If the remaining e-fold variable
\[
N_{\rm to\,end}=N_{\rm end}-N
\]
were adopted instead, an additional minus sign would appear in the derivative relation. Both the numerical implementation and the notation used here are formulated in terms of the forward variable \(N\). The sensitivity of the results to variations in \(\Delta N\) is included as part of the stability analysis of the primordial spectrum calculation.

The tensor spectrum is evaluated using Eq.~\eqref{eq:tensor}. Consequently, the observables \(A_s\), \(n_s\), and \(r_{0.05}\) are all determined self-consistently from the same warm inflation background trajectory, rather than being adjusted independently.

\subsection{Compressed Observable Likelihood and Prior-Averaged Weight}\label{sec:likelihood}

The calculation uses a constructed likelihood in the three pivot observables \(A_s\), \(n_s\), and \(r_{0.05}\), motivated by compressed-parameter approaches \cite{Prince:2019hse, Balkenhol:2024obe}. It is not the Planck or BICEP/Keck data likelihood itself. For a parameter set \(\theta\) and monomial power \(p\), the numerical trajectory gives \(H_*\), \(\dot{\phi}_*\), \(Q_*\), and \(T_*\), from which
\begin{equation}
\mu(\theta|p)=
\bigl(\ln(10^{10}A_s),\, n_s,\, r_{0.05}\bigr)
\end{equation}
is calculated. The three entries are therefore linked predictions of the same background and spectrum prescription.

The scalar amplitude and the scalar spectral index are assumed to follow Gaussian likelihood distributions:
\begin{align}
\chi^2_A &=
\left[
\frac{\ln(10^{10}A_s)-\ln(10^{10}A_s)_{\rm obs}}
{\sigma_A}
\right]^2,\\
\chi^2_n &=
\left[
\frac{n_s-n_{s,{\rm obs}}}{\sigma_n}
\right]^2 .
\end{align}

The numerical inputs are listed in Tab.~\ref{tab:compressed-likelihood}. Because the observational information on \(r\) is primarily an upper limit, it is represented by the one-sided function
\begin{equation}
\chi_r^2=\begin{cases}
0, & r_{0.05}\le r_{\rm lim},\\[2mm]
\left[(r_{0.05}-r_{\rm lim})/\sigma_r\right]^2, & r_{0.05}>r_{\rm lim}.
\end{cases}
\label{eq:rpenalty}
\end{equation}
Predictions below \(r_{\rm lim}\) receive no additional weight for being smaller, while predictions above the limit are penalized smoothly. The reference values are \(r_{\rm lim}=0.036\) and \(\sigma_r=0.005\). The choices \(\sigma_r=0.003,0.005,0.010\) are used to test whether this particular smoothing choice changes the ordering.

\begin{table}[htbp]
\centering
\caption{Inputs used in the constructed observable-space likelihood. The adopted scalar-amplitude width, \(\sigma_A=0.08\), is broader than that of the full Planck posterior, and \(r\) is represented by a one-sided upper-limit penalty.}
\label{tab:compressed-likelihood}
\begin{tabular}{ccc}   
\toprule
Observables & Adopted value or limit & Likelihood treatment \\
\midrule
$\ln(10^{10}A_s)$ & 3.044 & Gaussian, $\sigma_A=0.08$ \\
$n_s$ & 0.9649 & Gaussian, $\sigma_n=0.0042$ \\
$r_{0.05}$ & $<0.036$ & One-sided penalty, $\sigma_r=0.005$ \\
\bottomrule
\end{tabular}
\end{table}

The trajectory must also pass the specified feasibility checks. The total function is
\begin{equation}
\chi^2=\chi^2_A+\chi^2_n+\chi^2_r+
\chi^2_{\rm warm}+\chi^2_{\rm phys},
\qquad
\mathcal L_{\rm obs}=\exp\!\left(-\frac12\chi^2\right).
\label{eq:likelihood}
\end{equation}
Here \(\chi^2_{\rm warm}\) imposes \(T_*/H_*>1\), and \(\chi^2_{\rm phys}\) suppresses trajectories with negative radiation density, insufficient e-folds, failure of the low-temperature check, nonfinite observables, or no inflationary endpoint. Thus \(\mathcal L_{\rm obs}\) is a constructed pointwise weight combining the three observable terms and the adopted feasibility conditions. It is not by itself a model evidence.

The choice \(\sigma_A=0.08\) is part of the definition of the present calculation. It is broader than the uncertainty associated with the full Planck posterior. Adopting a narrower value, such as \(\sigma_A\simeq0.014\), would define a different likelihood and would require a new prior integration. The quoted \(Z_{\rm eff}\) values are therefore conditional on \(\sigma_A=0.08\); the broadening should not be interpreted as a full-CMB treatment or as a uniquely conservative statistical prescription.

For a given power \(p\), the prior-averaged weight is
\begin{equation}
\Geff(p)=
\int_{\Theta_p}
\mathcal L_{\rm obs}(\theta|p)\,\pi(\theta|p)\,\dd\theta,
\label{eq:zeff}
\end{equation}
where \(\Theta_p\) is the adopted prior domain and \(\pi(\theta|p)\) is the normalized prior density \cite{Martin:2024qnn, Easther:2011yq}. Equation~\eqref{eq:zeff} has the integral form of a Bayesian evidence, but its likelihood is the broadened observable-space function defined above. We retain the notation \(Z_{\rm eff}\) and use ``effective evidence'' only as shorthand for this prior-averaged constructed likelihood. It is conditional on both \(\mathcal L_{\rm obs}\) and the model-dependent prior domain and is not the evidence obtained from a full-CMB data likelihood. The integral accounts for both the pointwise likelihood and the fraction of the adopted prior domain that receives appreciable weight.

We report
\begin{equation}
\Delta\ln\Geff(p)=\ln\Geff(p)-\ln\Geff(p=4).
\end{equation}
A negative value means that the branch has a smaller prior-averaged weight than \(p=4\) within the same stated setup. The magnitude is not interpreted beyond the constructed likelihood and prior family.

\section{Numerical Strategy, Prior Construction, and \texorpdfstring{\(Z_{\rm eff}\)}{Zeff} Calculation}\label{sec:statistics}
\subsection{Computational Procedure and Division of Roles}\label{sec:calculation-flow}

The calculation is performed separately for \(p=2,3,4\) with the same background and spectrum code. Samples are drawn from the corresponding adopted prior domain, the full background system is integrated, and the pivot observables are evaluated. Trajectories with insufficient e-folds, negative radiation density, failure of the stated low-temperature or warm-background checks, or nonfinite observables receive a negligible likelihood. The same constructed likelihood in \((A_s,n_s,r_{0.05})\) is then applied to all three branches.

The genetic algorithm (GA) is used to locate viable high-likelihood regions and to select the representative trajectories in Tab.~\ref{tab:bestfit}. It does not determine \(Z_{\rm eff}\). The prior integral is evaluated with \texttt{dynesty}, and MCMC chains are used to check the posterior regions and sampling convergence. The additional variations of \(N_*\), the prior domains, random seeds, \(\sigma_r\), and the spectrum settings are finite sensitivity tests. They show whether the ordering changes for those choices; they do not establish invariance under arbitrary likelihoods, priors, or perturbation prescriptions.

\begin{table}[htbp]
\centering
\caption{Representative high-likelihood points used for trajectory and spectrum diagnostics. They are selected after broad exploration and GA localization and are reevaluated with the common background, spectrum, and constructed-likelihood pipeline. They are neither posterior means nor outputs of the \(Z_{\rm eff}\) integration.}
\label{tab:bestfit}
\begin{tabular}{ccccccc}
\toprule
Model & $n_s$ & $r_{0.05}$ & $A_s$ & $Q_*$ & $T_*/H_*$ & $\chi^2$ \\
\midrule
$p=2$ & 0.996616 & 0.047617 & $3.331\times10^{-9}$ & $6.19\times10^{-4}$ & 4.442 & 95.698 \\
$p=3$ & 0.977578 & 0.030739 & $2.306\times10^{-9}$ & $2.93\times10^{-3}$ & 8.509 & 10.482 \\
$p=4$ & 0.964201 & 0.026627 & $1.911\times10^{-9}$ & $4.68\times10^{-3}$ & 10.670 & 1.415 \\
\bottomrule
\end{tabular}
\end{table}

\subsection{Parameter Space and Viability-Conditioned Priors}

For each \(p\), the sampled parameters are
\begin{equation}
\theta=(\log_{10}\lambda_p,\log_{10}C_\phi,\phi_0,
\log_{10}f_{\dot\phi},\log_{10}f_{\rho_R}).
\label{eq:theta}
\end{equation}

The initial velocity and initial radiation density are written as
\begin{equation}
\dot\phi_0=10^{\log_{10}f_{\dot\phi}}\dot\phi_{\rm sr},\qquad
\rho_{R,0}=10^{\log_{10}f_{\rho_R}}V(\phi_0),
\end{equation}
where \(\dot\phi_{\rm sr}=-V_{,\phi}/(3H+\Upsilon)\). The reference prior A is shown in Tab.~\ref{tab:priorA}; the two additional prior sets used for sensitivity testing are given in Appendix~\ref{app:priorboxes}.

\begin{table}[htbp]
\centering
\caption{Reference prior A. The distributions are uniform in the listed coordinates. The model-dependent ranges were selected after exploratory scans identified the principal viable branch for each \(p\); the resulting \(Z_{\rm eff}\) values are conditional on this construction.}
\label{tab:priorA}
\begin{tabular}{cccccc}
\toprule
Model & \(\log_{10}\lambda_p\) & \(\log_{10}C_\phi\) & \(\phi_0\) & \(\log_{10}f_{\dot\phi}\) & \(\log_{10}f_{\rho_R}\) \\
\midrule
\(p=2\) & [-12.30,-9.20] & [5.00,9.40] & [12.0,38.0] & [-1.50,0.10] & [-14.0,-7.0] \\
\(p=3\) & [-13.30,-10.00] & [5.60,9.60] & [18.0,42.0] & [-1.40,0.10] & [-14.0,-7.0] \\
\(p=4\) & [-13.90,-12.60] & [6.40,8.70] & [30.0,42.0] & [-1.50,-0.30] & [-14.0,-8.5] \\
\bottomrule
\end{tabular}
\end{table}

Prior A is defined after exploratory scans identify the principal viable branch for each monomial power. The scans use the requirements of sufficient e-folds, \(T_*/H_*>1\), positive radiation density, finite observables, and non-negligible constructed likelihood. The resulting domains are therefore model-dependent and viability-conditioned; they are not independent of the exploratory calculation. Priors B and C, listed in Appendix~\ref{app:priorboxes}, alter selected widths and boundaries to test the dependence of the ordering on this finite family of prior choices. The GA locates candidate regions but does not generate posterior samples or evaluate the integral in Eq.~\eqref{eq:zeff}.

Identical coordinate widths are not imposed across the three powers because \(\lambda_p\) has a \(p\)-dependent mass dimension and the viable branches occupy different numerical scales in the chosen coordinates. This choice does not make the priors unique or objectively fair. It defines one model-dependent comparison, whose sensitivity is probed only by priors A--C. A prior specified in common pivot-scale physical variables could lead to different numerical evidence values and would require a separate calculation.

\begin{table}[htbp]
\centering
\caption{Prior box width diagnostics for reference prior A. \(\ln V_{\rm box}\) is only used to display the difference in prior ranges; the evidence integration uses the normalized prior density.}
\label{tab:prior volume-A}
\begin{tabular}{ccccccc}
\toprule
Model & $\Delta\log_{10}\lambda_p$ & $\Delta\log_{10}C_\phi$ & $\Delta\phi_0$ & $\Delta\log_{10}f_{\dot\phi}$ & $\Delta\log_{10}f_{\rho_R}$ & $\ln V_{\rm box}$ \\
\midrule
$p=2$ & 3.10 & 4.40 & 26.0 & 1.60 & 7.00 & 8.287 \\
$p=3$ & 3.30 & 4.00 & 24.0 & 1.50 & 7.00 & 8.110 \\
$p=4$ & 1.30 & 2.30 & 12.0 & 1.20 & 5.50 & 5.467 \\
\bottomrule
\end{tabular}
\end{table}

Tab.~\ref{tab:prior volume-A} displays the coordinate widths of prior A; the integral itself uses normalized prior densities. Priors A--C test whether the ordering reverses under the particular domain changes specified in Appendix~\ref{app:priorboxes}. The ordering does not reverse, but the magnitude of the differences is prior-dependent: for example, \(\Delta\ln Z_{\rm eff}(p=2)\) changes from \(-32.18\) under prior A to \(-11.36\) under prior C. The sensitivity test therefore supports only the limited statement that the ordering is retained for these three prior definitions.

\subsection{Global Search, Posterior Diagnostics, and Prior Integration}

The nonlinear background equations produce many invalid trajectories and comparatively narrow viable regions. A GA is therefore used as a gradient-free search tool to locate viable high-likelihood regions \cite{Cheng:2025rmf, Bernardo:2025zbv}. The search is restricted to the adopted prior domains and is not used to calculate \(Z_{\rm eff}\) or to redefine those domains during nested sampling.

Each sample is a five-dimensional vector of Eq.~\eqref{eq:theta}. The prior transform maps the unit hypercube to the fixed intervals in Tab.~\ref{tab:priorA} or Appendix~\ref{app:priorboxes}. For every sample, the background and observables are recomputed. Invalid trajectories receive a negligible log-likelihood within the same fixed domain. Thus the reported \(Z_{\rm eff}\) values are integrals over the stated domains, not local integrals around GA solutions.

Posterior samples are obtained with the affine-invariant ensemble MCMC method \cite{Allison:2013npa, Foreman-Mackey:2012any}, and \(Z_{\rm eff}\) is evaluated with \texttt{dynesty}. The reference nested-sampling runs use \(N_{\rm live}=500\), \texttt{bound=multi}, \texttt{sample=rwalk}, and the stopping criterion \(d\ln Z=0.05\). MCMC and nested-sampling posterior samples are generated from the same prior domains and likelihood function.

The internal \texttt{dynesty} uncertainty is reported separately from the scatter among three random seeds. A normalizing-flow reconstruction is used only to compare the location of the principal quartic posterior region with the MCMC and weighted nested-sampling samples. It does not enter the \(Z_{\rm eff}\) calculation. The three seeds give the same ordering for the tested configuration; this establishes neither seed-independent exactness nor robustness to changes outside the reported setup.

\begin{table}[htbp]
\centering
\caption{Long-chain MCMC diagnostics. Here \(\tau_{\rm int}\) is the integrated autocorrelation time, ESS is the effective sample size, and split-\(\hat R\) measures between-chain agreement.}
\label{tab:mcmc}
\resizebox{\textwidth}{!}{%
\begin{tabular}{ccccccc}
\toprule
Model & Number of walkers & Number of steps & Mean acceptance fraction & \(\max\tau_{\rm int}\) & Minimum ESS & Maximum split-\(\hat R\) \\
\midrule
\(p=2\) & 80 & 25000 & 0.405 & 114.3 & 17500 & 1.0056 \\
\(p=3\) & 80 & 25000 & 0.398 & 116.1 & 17228 & 1.0050 \\
\(p=4\) & 80 & 25000 & 0.382 & 125.5 & 15934 & 1.0057 \\
\bottomrule
\end{tabular}%
}
\end{table}

\section{Numerical Results and Sensitivity Tests}\label{sec:results}

This section reports the representative trajectories, the \(Z_{\rm eff}\) values, and the finite sensitivity tests. The comparison covers the location of the three branches in the \(n_s-r_{0.05}\) plane, the changes obtained for \(N_*=50,55,60\), priors A--C, three random seeds, three values of \(\sigma_r\), selected spectrum settings, and the background validity diagnostics. Unless stated otherwise, \(\sigma_r=0.005\) is used. Throughout this section, \(Z_{\rm eff}\) denotes the integral of the constructed observable-space likelihood over the adopted viability-conditioned prior domain.

\subsection{Representative High-Likelihood Trajectories in the \texorpdfstring{\(n_s-r_{0.05}\)}{ns-r0.05} Plane}\label{sec:bestfits}

Tab.~\ref{tab:bestfit} lists one representative high-likelihood trajectory for each power. The points are selected after exploratory scans and GA localization and are reevaluated with the common pipeline. They are used only for trajectory diagnostics, spectrum-component tests, and illustration in the \(n_s-r_{0.05}\) plane. They are neither posterior means nor substitutes for the prior integral in Eq.~\eqref{eq:zeff}.

At the listed points, \(p=2\) gives \(n_s=0.996616\) and \(r_{0.05}=0.047617\), producing the largest \(\chi^2\). The cubic point lies below the adopted tensor limit but has \(n_s=0.977578\). The quartic point has \(n_s=0.964201\) and \(r_{0.05}=0.026627\). These values show that the warm-spectrum correction does not shift the three powers by a common amount.

\begin{figure}[htbp]
\centering
\includegraphics[width=0.80\textwidth, height=0.75\textheight, keepaspectratio]{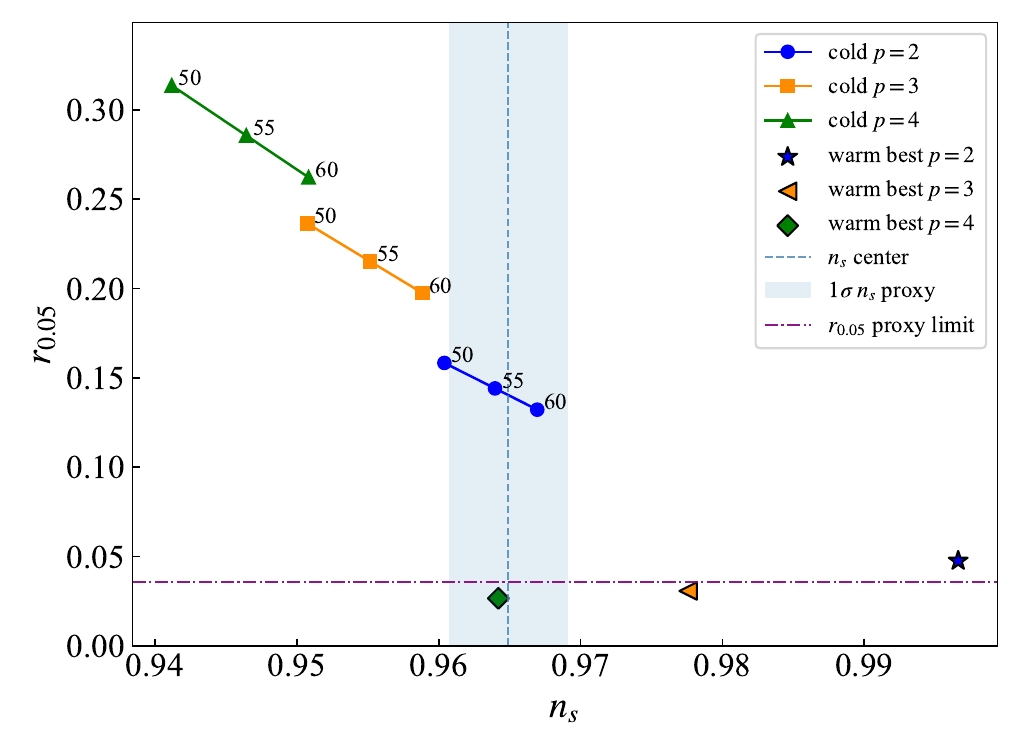}
\caption{Analytic cold-inflation reference points and representative warm trajectories in the \(n_s-r_{0.05}\) plane. The shaded band and horizontal dashed line show the \(n_s\) interval and \(r_{0.05}=0.036\) limit used in the constructed likelihood. The cold points are not included in the \(Z_{\rm eff}\) calculation.}
\label{fig:nsr}
\end{figure}

Fig.~\ref{fig:nsr} compares the analytic cold reference points with the representative warm trajectories. The cold points are not included in the \(Z_{\rm eff}\) calculation. Under the constructed likelihood, the representative warm points lie at smaller \(r\), and the quartic point falls below the adopted limit \(r_{0.05}=0.036\).

At the representative-point level, the quadratic trajectory is farthest from the adopted observable targets, the cubic trajectory is intermediate, and the quartic trajectory has the smallest \(\chi^2\). This pointwise comparison does not determine how much of each adopted prior domain has appreciable likelihood. The next subsection therefore reports the prior integrals.

\subsection{Conditional \texorpdfstring{\(Z_{\rm eff}\)}{Zeff} Ordering and Sensitivity Tests}\label{sec:evidence-results}

Tab.~\ref{tab:evidence} summarizes the effective evidence results obtained for \(N_*=55\), \(\sigma_r=0.005\), and reference prior A. The quantity \(\ln Z_{\rm eff}\) is evaluated using Eq.~\eqref{eq:zeff} and corresponds to the integral of the effective observable likelihood over the adopted prior region, while \(\Delta\ln Z_{\rm eff}\) denotes the evidence difference measured relative to the quartic potential.

\begin{table}[htbp]
\centering
\caption{Effective evidence under \(N_*=55\) and reference prior A. \(N_{\rm live}=500\); the numerical integration error is the internal uncertainty of \(\ln Z_{\rm eff}\) from a single nested sampling run.}
\label{tab:evidence}
\begin{tabular}{ccccc}
\toprule
Model & $N_{\rm live}$ & $\ln Z_{\rm eff}$ & Numerical integration error & $\Delta\ln Z_{\rm eff}$ \\
\midrule
$p=2$ & 500 & -37.1116 & 0.1325 & -32.1769 \\
$p=3$ & 500 & -11.9207 & 0.1288 & -6.9861 \\
$p=4$ & 500 & -4.9347 & 0.1048 & 0 \\
\bottomrule
\end{tabular}
\end{table}

For the reference setup, Tab.~\ref{tab:evidence} gives the ordering \(p=4>p=3\gg p=2\) in \(Z_{\rm eff}\). The differences are \(-6.99\) for \(p=3\) and \(-32.18\) for \(p=2\), relative to \(p=4\). These differences exceed the internal nested-sampling errors of approximately \(0.1\). They quantify the prior-averaged constructed likelihood within prior A; they are not Bayes factors from a full-CMB likelihood.

Tab.~\ref{tab:nstar} repeats the calculation for \(N_*=50,55,60\). The ordering does not change for these three choices, although the numerical difference for the cubic branch varies from \(-5.01\) to \(-8.09\).

\begin{table}[htbp]
\centering
\caption{$N_*$ discrete sensitivity test. The quartic potential serves as the zero point reference.}
\label{tab:nstar}
\begin{tabular}{cccc}
\toprule
$N_*$ & $\Delta\ln Z_{\rm eff}(p=2)$ & $\Delta\ln Z_{\rm eff}(p=3)$ & $\ln Z_{\rm eff}(p=4)$ \\
\midrule
50 & -31.716 & -5.005 & -5.247 \\
55 & -32.177 & -6.986 & -4.935 \\
60 & -32.334 & -8.090 & -5.320 \\
\bottomrule
\end{tabular}
\end{table}

Tab.~\ref{tab:prior-sens} shows the dependence on priors A--C. The ordering is unchanged, but the differences are not stable in magnitude. In particular, \(\Delta\ln Z_{\rm eff}(p=2)\) changes from \(-32.18\) for prior A to \(-11.36\) for prior C. The table therefore supports an ordering that is retained for these three domains, not prior-independent evidence strengths.

\begin{table}[htbp]
\centering
\caption{Sensitivity to the three prior domains defined in Appendix~\ref{app:priorboxes}. Prior A is the reference viability-conditioned domain; priors B and C alter selected widths and boundaries. The table tests only these specified prior choices.}
\label{tab:prior-sens}
\begin{tabular}{cccc}
\toprule
Prior set & $\Delta\ln Z_{\rm eff}(p=2)$ & $\Delta\ln Z_{\rm eff}(p=3)$ & $\ln Z_{\rm eff}(p=4)$ \\
\midrule
A:Reference prior & -32.177 & -6.986 & -4.935 \\
B:Unified auxiliary parameter widths & -31.628 & -6.247 & -5.665 \\
C:Extended prior ranges & -11.361 & -7.041 & -6.371 \\
\bottomrule
\end{tabular}
\end{table}

Tab.~\ref{tab:seed} reports three independent random seeds for the reference configuration. The ordering is the same in all three runs. The cubic difference lies between \(-6.99\) and \(-6.74\), and the quadratic difference remains close to \(-32.1\). This checks the reported seed scatter but does not exclude every form of numerical systematic error.

\begin{table}[htbp]
\centering
\caption{Results for three independent random seeds at \(N_*=55\), reported relative to the quartic branch.}
\label{tab:seed}
\begin{tabular}{cccc}
\toprule
Random seed & $\Delta\ln Z_{\rm eff}(p=2)$ & $\Delta\ln Z_{\rm eff}(p=3)$ & $\ln Z_{\rm eff}(p=4)$ \\
\midrule
11 & -32.177 & -6.986 & -4.935 \\
22 & -32.080 & -6.761 & -5.050 \\
33 & -32.105 & -6.744 & -5.092 \\
\bottomrule
\end{tabular}
\end{table}

Tab.~\ref{tab:sigmar} varies the smoothing width of the one-sided \(r\) penalty over \(\sigma_r=0.003,0.005,0.010\). The ordering is unchanged for these values, and the cubic difference remains between \(-6.84\) and \(-6.99\). This test addresses only the adopted penalty form and the three widths shown.

\begin{table}[htbp]
\centering
\caption{Sensitivity to the smoothing width \(\sigma_r\) of the one-sided \(r\) penalty at \(N_*=55\). Differences are reported relative to the quartic branch.}
\label{tab:sigmar}
\begin{tabular}{cccc}
\toprule
$\sigma_r$ & $\Delta\ln Z_{\rm eff}(p=2)$ & $\Delta\ln Z_{\rm eff}(p=3)$ & $\ln Z_{\rm eff}(p=4)$ \\
\midrule
0.003 & -32.698 & -6.840 & -4.985 \\
0.005 & -32.177 & -6.986 & -4.935 \\
0.010 & -31.637 & -6.889 & -4.818 \\
\bottomrule
\end{tabular}
\end{table}

Taken together, Tabs.~\ref{tab:evidence}--\ref{tab:sigmar} show that the ordering \(p=4>p=3>p=2\) is retained for the reported choices of \(N_*\), prior domain, random seed, and \(\sigma_r\). The tables do not test a full-CMB likelihood, arbitrary prior constructions, or alternative inflaton occupation models.

\subsection{Spectrum-Calculation Checks and Low-Temperature Trajectory Diagnostics}\label{sec:spectrum-results}

Tab.~\ref{tab:spectrum} separates two types of checks. The first varies the finite-difference step and the fitted growth factor by \(\pm10\%\); these are routine numerical or fitting-prescription variations. The second sets \(n_*=0\) at the representative trajectories. This changes the assumed perturbation state and is included only to isolate the occupation contribution. It is neither a numerical-error estimate nor a recomputation of \(Z_{\rm eff}\).

\begin{table}[htbp]
\centering
\caption{Spectrum checks at the representative trajectories. ``Maximum routine variation'' combines changes in the finite-difference step and \(\pm10\%\) changes in \(G(Q)\). ``Occupation number removed'' gives the change obtained by setting \(n_*=0\); it is a component test of the assumed Bose--Einstein occupation, not a numerical uncertainty or a new \(Z_{\rm eff}\) calculation.}
\label{tab:spectrum}
\begin{tabular}{ccccccc}
\toprule
 & \multicolumn{3}{c}{Maximum routine variation} & \multicolumn{3}{c}{Occupation number removed} \\
Model & $|\Delta n_s|$ & $|\Delta r|$ & $|\Delta\ln(10^{10}A_s)|$ & $|\Delta n_s|$ & $|\Delta r|$ & $|\Delta\ln(10^{10}A_s)|$ \\
\midrule
$p=2$ & $3.81\times10^{-7}$ & $1.35\times10^{-8}$ & $2.84\times10^{-7}$ & 0.098 & 0.371 & 2.173 \\
$p=3$ & $2.76\times10^{-6}$ & $1.80\times10^{-7}$ & $5.86\times10^{-6}$ & 0.0458 & 0.427 & 2.700 \\
$p=4$ & $5.13\times10^{-6}$ & $3.88\times10^{-7}$ & $1.46\times10^{-5}$ & 0.0155 & 0.413 & 2.805 \\
\bottomrule
\end{tabular}
\end{table}

The routine variations change \(n_s\), \(r\), and \(\ln(10^{10}A_s)\) by much less than the separations among the three listed trajectories. Setting \(n_*=0\), by contrast, raises \(r\) to values of order \(0.4\) at those points. The component test therefore confirms that the baseline predictions of the representative trajectories depend strongly on the assumed occupation term, whereas the fitted \(G(Q)\) correction is small in the weak-dissipation CMB window. Because the prior integration is not repeated, this test does not establish the \(Z_{\rm eff}\) ordering for \(n_*=0\) or for partial thermal occupation.

\begin{figure}[htbp]
\centering
\includegraphics[width=0.80\textwidth, height=0.75\textheight, keepaspectratio]{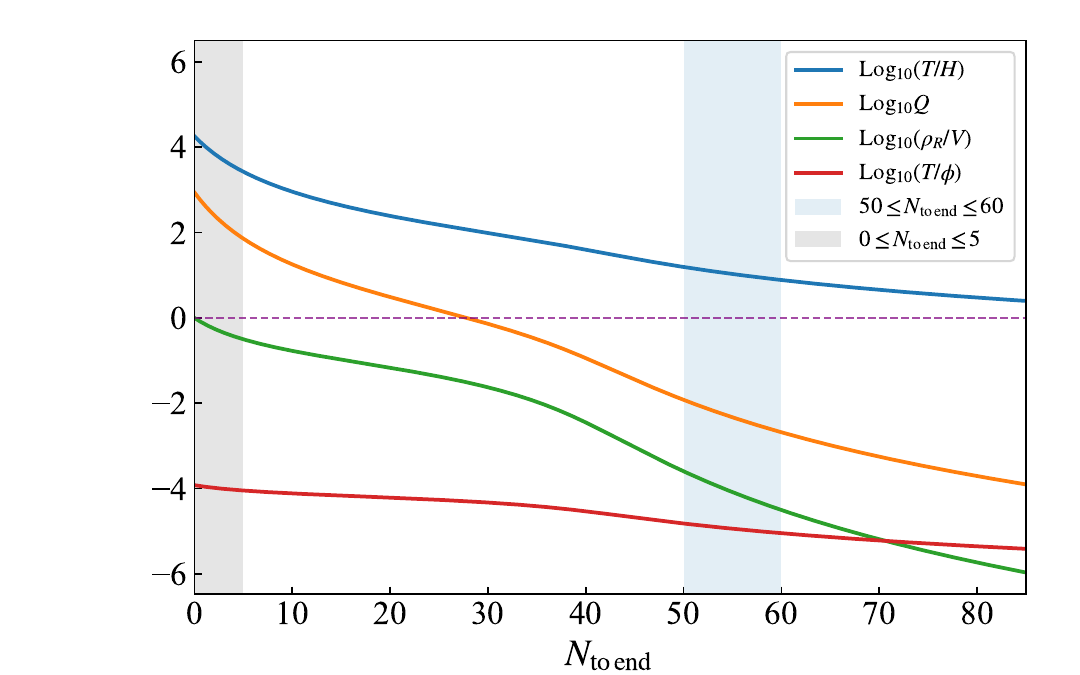}
\caption{Validity diagnostics for the representative trajectory of the quartic potential. The horizontal axis shows the number of e-folds remaining until the end of inflation, \(N_{\rm to\,end}\); the light blue shaded region for \(50\le N_{\rm to\,end}\le60\) indicates the CMB perturbation generation window, the grey shaded region for \(0\le N_{\rm to\,end}\le5\) indicates the late-time radiation-enhanced stage, and the dotted vertical line marks the reference pivot position.}
\label{fig:p4traj}
\end{figure}

\begin{figure}[htbp]
\centering
\includegraphics[width=0.80\textwidth, height=0.75\textheight, keepaspectratio]{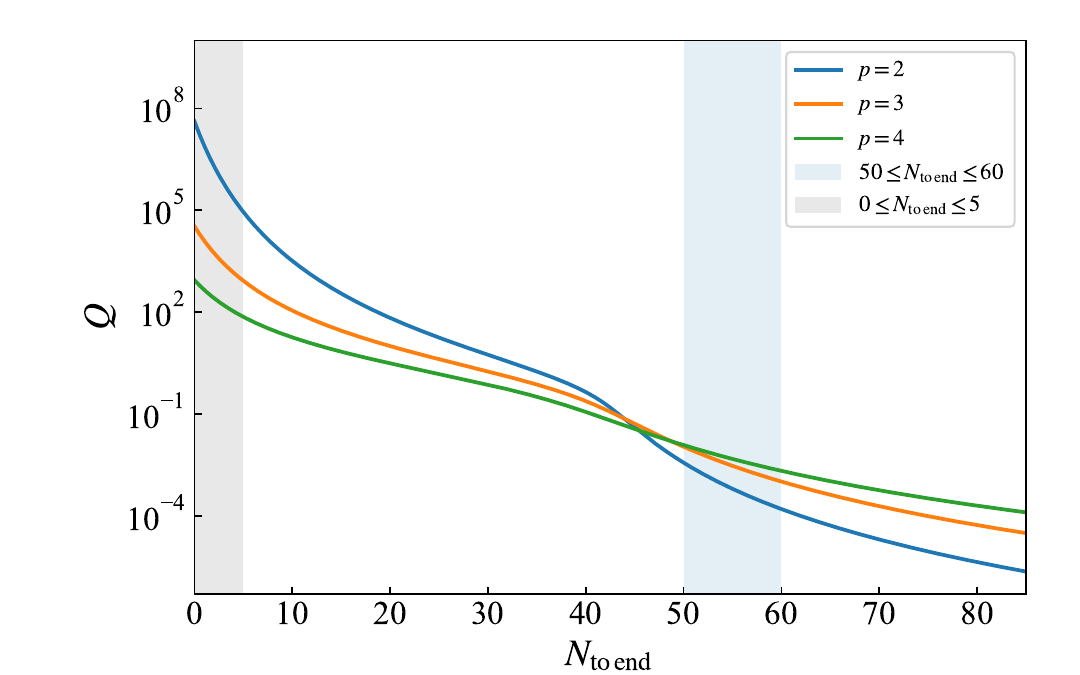}
\caption{Evolution of the dissipation ratio \(Q\) for the three representative trajectories. The light blue shaded region indicates the CMB perturbation generation window, and the grey shaded region indicates the late-time radiation-enhanced stage. In the CMB window \(Q\ll1\), and both radiation and dissipation increase rapidly in the last few e-folds.}
\label{fig:qtraj}
\end{figure}

Trajectory-level diagnostics are reported in Tab.~\ref{tab:micro} and Figs.~\ref{fig:p4traj} and \ref{fig:qtraj}. The CMB interval is defined by \(50\le N_{\rm to\,end}\le60\); the final five e-folds are shown separately to distinguish the perturbation-generation interval from the exit stage.

\begin{table}[htbp]
\centering
\caption{Windowed background diagnostics. The CMB interval is \(50\le N_{\rm to\,end}\le60\). The final five e-folds are shown only to display the radiation-enhanced exit stage and are not used as a CMB-window exclusion criterion.}
\label{tab:micro}
\resizebox{\textwidth}{!}{%
\begin{tabular}{ccccccc}
\toprule
Model & Window & \(T/H\) range & $\max Q$ & $\max(\rho_R/V)$ & $\max(T/\phi)$ & $\max\epsilon_H$ \\
\midrule
$p=2$ & pivot & 4.42--4.42 & $6.08\times10^{-4}$ & $7.9\times10^{-6}$ & $1.4\times10^{-5}$ & 0.027 \\
$p=2$ & CMB window & 2.82--7.87 & $3.45\times10^{-3}$ & $5.7\times10^{-5}$ & $2.6\times10^{-5}$ & 0.036 \\
$p=3$ & pivot & 8.55--8.55 & $2.97\times10^{-3}$ & $4.7\times10^{-5}$ & $1.4\times10^{-5}$ & 0.033 \\
$p=3$ & CMB window & 5.84--13.58 & $1.05\times10^{-2}$ & $2.1\times10^{-4}$ & $2.1\times10^{-5}$ & 0.042 \\
$p=4$ & pivot & 10.71--10.71 & $4.73\times10^{-3}$ & $8.4\times10^{-5}$ & $1.2\times10^{-5}$ & 0.036 \\
$p=4$ & CMB window & 7.72--15.48 & $1.18\times10^{-2}$ & $2.5\times10^{-4}$ & $1.5\times10^{-5}$ & 0.044 \\
$p=4$ & Last 5 e-folds & $2.77\times10^3$--$1.81\times10^4$ & $8.71\times10^2$ & 0.999 & $1.2\times10^{-4}$ & 1.000 \\
\bottomrule
\end{tabular}}
\end{table}

For the representative trajectories, \(T/H>1\) throughout the CMB interval, while \(Q\) and \(\rho_R/V\) remain small. For \(p=4\), the largest reported CMB-window values are \(Q\simeq1.18\times10^{-2}\) and \(\rho_R/V\simeq2.5\times10^{-4}\). If \(m_\chi=g_\chi\phi\), the tabulated \(T/\phi\sim10^{-5}\) gives the lower bound \(g_\chi>T/\phi\). Together with \(T/H\sim10\), this permits \(m_\chi/H\gg1\) for a range of \(g_\chi\). These inequalities are consistency checks on the effective trajectories; they do not identify a complete mediator sector.

During the final few e-folds, \(\rho_R/V\) rises toward unity. This interval is not used to evaluate the pivot observables. The reported feasibility criterion requires radiation subdominance in the CMB interval and at the pivot, while allowing a radiation-enhanced stage near \(\epsilon_H=1\).

\begin{figure}[htbp]
\centering
\includegraphics[width=0.80\textwidth, height=0.75\textheight, keepaspectratio]{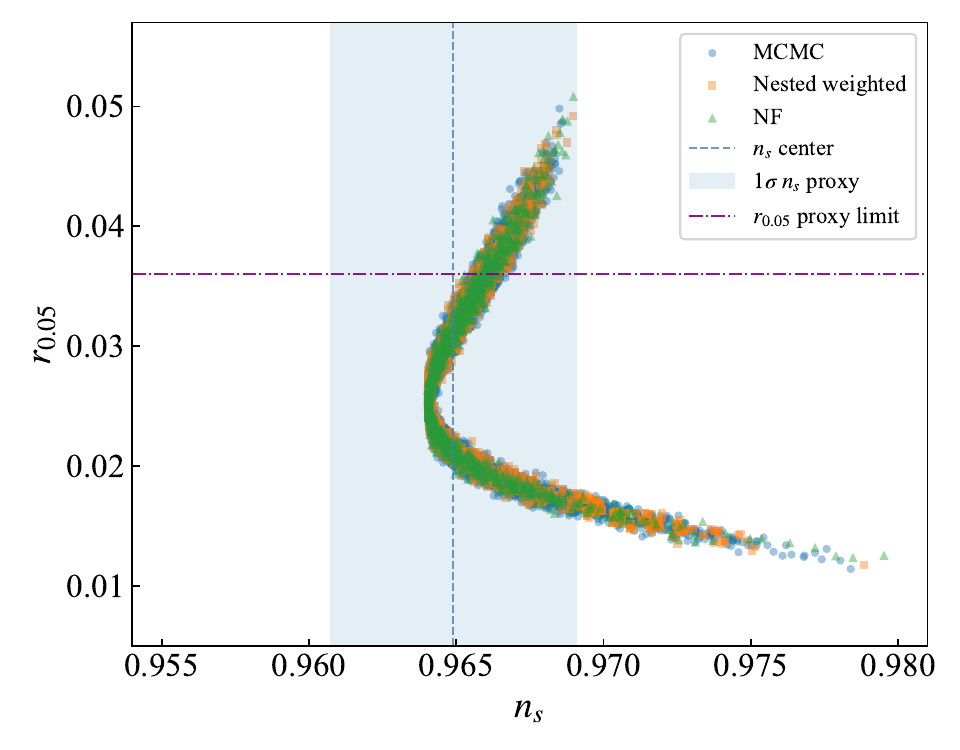}
\caption{Quartic-branch samples in the \(n_s-r_{0.05}\) plane from MCMC, weighted \texttt{dynesty} output, and a normalizing-flow reconstruction. The dashed lines show \(n_s=0.9649\) and \(r_{0.05}=0.036\). The comparison checks the location of the principal sampled region; it is not used to calculate or validate \(Z_{\rm eff}\).}
\label{fig:methodoverlap}
\end{figure}

The three sample sets overlap in the principal quartic region shown in Fig.~\ref{fig:methodoverlap}. Together with \(\hat R<1.01\), chain lengths exceeding \(50\tau_{\rm int}\), and \(\mathrm{ESS}>1.5\times10^4\), this provides no indication of non-convergence in the reported quartic posterior calculation. It does not test the physical assumptions entering the spectrum or prior construction.

Within the constructed likelihood and the adopted prior family, \(p=4\) has the largest \(Z_{\rm eff}\), \(p=3\) is intermediate, and \(p=2\) has the smallest value. This statement is conditional on the fixed dissipative coefficient, the Bose--Einstein occupation prescription, and the model-dependent prior domains. The required values \(C_\phi\sim10^6\)--\(10^7\) are effective dissipative strengths and remain a nontrivial microscopic requirement.

\section{Physical Discussion}\label{sec:discussion}
\subsection{Occupation Dependence of the Representative Quartic Trajectory}

For the representative quartic trajectory, \(Q_*\simeq4.7\times10^{-3}\), so \(G(Q_*)\) is close to unity. Under the adopted Bose--Einstein prescription, the larger contribution to the scalar enhancement comes from \(1+2n_*\). This lowers \(r\) relative to the same trajectory evaluated without the occupation term. The role of thermal occupation in quartic warm inflation is established in earlier studies \cite{Bartrum:2013fia, Ballesteros:2023dno}; the present component test quantifies it for the representative trajectory used in this comparison.

If the actual inflaton occupation lies below the Bose--Einstein value, the scalar enhancement and the reduction in \(r\) are smaller. The \(n_*=0\) entries in Tab.~\ref{tab:spectrum} illustrate the limiting change at the representative trajectories. Because \(Z_{\rm eff}\) is not recomputed, no evidence ordering is claimed for \(n_*=0\), partial occupation, or a specified microscopic thermalization model. The baseline ordering applies only under the occupation prescription of Eq.~\eqref{eq:nBE}.

\subsection{Interpretation of \texorpdfstring{\(Z_{\rm eff}\)}{Zeff}}

The quantity \(Z_{\rm eff}^{(A_s,n_s,r)}\) is the normalized prior average of the constructed observable-space likelihood. It is useful for applying the same numerical criterion to the three monomial powers, but it is not the evidence associated with the Planck and BICEP/Keck likelihoods. Its value depends on the adopted observable widths, the upper-limit penalty, the feasibility terms, and the prior domain.

Within the present definition, \(p=4\) has the largest \(Z_{\rm eff}\) among \(p=2,3,4\), and the ordering is retained for the finite variations reported in Sec.~\ref{sec:results}. This result should be read as a conditional cross-potential ordering. It does not show that warm inflation generally selects a quartic potential, nor does it supersede full-CMB analyses of individual warm-inflation branches. The representative-point spectrum test indicates why the adopted quartic trajectory has a smaller \(r\), but it does not by itself explain the full prior-integrated difference.

\subsection{Prior Dependence and Microphysical Requirements}

Prior A was selected after exploratory scans identified a viable branch for each \(p\). It is therefore model-dependent and viability-conditioned. Priors B and C test selected changes in coordinate widths and boundaries; they do not cover all plausible prior constructions. The unchanged ordering across A--C is a useful internal sensitivity result, but the variation in \(\Delta\ln Z_{\rm eff}\), especially for \(p=2\), shows that the evidence strength is prior-dependent. A comparison based on common pivot-scale quantities or a hierarchical microphysical prior would be a distinct calculation.

The values \(C_\phi\sim10^6\)--\(10^7\) are a substantial model-building requirement. In a two-stage construction, \(C_\phi\) may receive multiplicity and coupling enhancements, schematically \(C_\phi\sim\mathcal C h^4N_\chi N_y^2\), but the numerical coefficient and multiplicities are model-dependent. The benchmark \(g_*=228.75\) counts relativistic bath degrees of freedom and should not be identified with \(C_\phi\). No explicit particle realization producing the required coefficient while satisfying all perturbative and thermal conditions is constructed in this paper.

\section{Conclusion}\label{sec:conclusion}

We compared the monomial potentials \(V_p(\phi)=\lambda_p\phi^p/p\), with \(p=2,3,4\), while fixing the dissipative coefficient to \(\Upsilon=C_\phi T^3/\phi^2\). The same background equations, primordial-spectrum prescription, and constructed likelihood in \((A_s,n_s,r_{0.05})\) were used for all three powers. This is a cross-potential comparison within an established warm-inflation setup; it does not introduce a new dissipative mechanism or a new explanation of thermal scalar enhancement.

For \(N_*=55\), \(\sigma_r=0.005\), and prior A, the differences relative to the quartic branch are \(\Delta\ln Z_{\rm eff}(p=2)=-32.18\) and \(\Delta\ln Z_{\rm eff}(p=3)=-6.99\). The same ordering, \(p=4>p=3\gg p=2\), is obtained for the selected values of \(N_*\), the three prior domains, the three random seeds, and the three upper-bound smoothing widths. The magnitude of the differences is not invariant: under prior C, the quadratic difference changes from \(-32.18\) to \(-11.36\). The ordering is therefore conditional on the constructed likelihood and the tested prior family.

The representative quartic trajectory has \(Q_*=4.68\times10^{-3}\) and \(T_*/H_*=10.67\). Under the assumed Bose--Einstein occupation prescription, the occupation term contributes more to the scalar enhancement than the fitted growth factor \(G(Q)\). Setting \(n_*=0\) at the representative points raises \(r\) to values of order \(0.4\). This component test shows the importance of the assumed occupation state for those trajectories, but it is not a new prior integration and does not establish the ordering for partially thermal or nonthermal inflaton fluctuations.

Across the reported CMB intervals, the representative trajectories satisfy \(T/H>1\), \(Q\ll1\), radiation subdominance, and values of \(T/\phi\) compatible with \(T<m_\chi\) and \(m_\chi\gg H\) for a range of \(g_\chi\). Near the end of inflation, the radiation fraction grows rapidly; this stage is displayed separately from the interval used to evaluate the pivot observables. The large coefficient \(C_\phi\sim10^6\)--\(10^7\) remains a substantial microphysical requirement, and no explicit realization is provided here.

Under the fixed low-temperature dissipative form, broadened observable-space likelihood, viability-conditioned priors, and the assumed Bose–Einstein inflaton occupation, the quartic branch carries the largest prior-averaged weight among \(p=2,3,4\). This confirmed ordering quantitatively distinguishes the statistical superiority of polynomial potentials within warm Higgs inflation, and furnishes concrete reference for potential selection and structural optimisation in warm inflation phenomenological research.

\section*{Acknowledgements}

This work was supported in part by the National Natural Science Foundation of China under Grant No. 12565014, by the Talent Research Startup Foundation of Hainan Normal University: HSZK-KYQD-202523, by Opening Foundation of Shanghai Key Laboratory of Particle Physics and Cosmology under Grant No. 22DZ2229013-5, by Hainan Provincial Natural Science Foundation of China under Grant No. 126MS0134, by Chongqing Natural Science Foundation project under Grant No. CSTB2022NSCQ-MSX0432, by Science and Technology Research Project of Chongqing Education Commission under Grant No. KJQN202200621 and No. KJQN202200650, and by Chongqing Human Resources and Social Security Administration Program under Grants No. D63012022005.

\appendix
\section{Viability-Conditioned Prior Ranges and Sensitivity Settings}\label{app:priorboxes}

This appendix lists the three prior domains A, B, and C. Prior A was defined after exploratory parameter scans, feasibility screening, and GA localization identified the principal viable branch for each monomial power. It is therefore a model-dependent, viability-conditioned prior. Prior B changes selected auxiliary-coordinate widths, while prior C extends the boundaries of prior A. These domains test whether the ordering changes under the specified variations; they do not establish invariance under arbitrary priors. All MCMC and nested-sampling calculations use the fixed domains listed below. The GA is used only to locate candidate regions and does not generate posterior samples or evaluate \(Z_{\rm eff}\).

\begin{table}[H]
\centering
\caption{Priors A, B, and C. All entries are uniform intervals for the corresponding variables.}
\label{tab:allpriors}
\resizebox{\textwidth}{!}{%
\begin{tabular}{ccccccc}
\toprule
Prior set & Model & \(\log_{10}\lambda_p\) & \(\log_{10}C_\phi\) & \(\phi_0\) & \(\log_{10}f_{\dot\phi}\) & \(\log_{10}f_{\rho_R}\) \\
\midrule
A & \(p=2\) & [-12.30,-9.20] & [5.00,9.40] & [12.0,38.0] & [-1.50,0.10] & [-14.0,-7.0] \\
A & \(p=3\) & [-13.30,-10.00] & [5.60,9.60] & [18.0,42.0] & [-1.40,0.10] & [-14.0,-7.0] \\
A & \(p=4\) & [-13.90,-12.60] & [6.40,8.70] & [30.0,42.0] & [-1.50,-0.30] & [-14.0,-8.5] \\
B & \(p=2\) & [-12.30,-9.20] & [5.00,9.40] & [12.0,38.0] & [-1.50,0.10] & [-14.0,-7.0] \\
B & \(p=3\) & [-13.30,-10.00] & [5.40,9.80] & [17.0,43.0] & [-1.45,0.15] & [-14.0,-7.0] \\
B & \(p=4\) & [-13.90,-12.60] & [5.35,9.75] & [23.0,49.0] & [-1.70,-0.10] & [-14.75,-7.75] \\
C & \(p=2\) & [-13.85,-7.65] & [2.80,11.60] & [1.0,51.0] & [-2.30,0.90] & [-17.5,-3.5] \\
C & \(p=3\) & [-14.95,-8.35] & [3.60,11.60] & [6.0,54.0] & [-2.15,0.85] & [-17.5,-3.5] \\
C & \(p=4\) & [-14.55,-11.95] & [5.25,9.85] & [24.0,48.0] & [-2.10,0.30] & [-16.75,-5.75] \\
\bottomrule
\end{tabular}}
\end{table}

\end{document}